\documentclass[%
 preprint,
 amsmath,amssymb,
 aps,
pra,
]{revtex4-2}
\usepackage{braket}
\usepackage{comment}
\usepackage{graphicx}
\usepackage{soul}
\usepackage{dcolumn}
\usepackage{bm}
\usepackage{hyperref}
\usepackage[normalem]{ulem}
\hypersetup{
    unicode=false,          
    pdftoolbar=false,        
    pdfmenubar=true,        
    pdffitwindow=false,     
    pdfstartview={FitH},    
    pdfkeywords={Quantum batteries} {Ergotropy} {Micromasers}{Steady State}, 
    pdfnewwindow=true,      
    colorlinks=true,       
    linkcolor=black,          
    citecolor=blue,        
    filecolor=magenta,      
    urlcolor=blue           
}

\usepackage{xcolor}
\usepackage{mathrsfs}
\usepackage{amsfonts}
\usepackage{amsmath}
\usepackage{amssymb}
\usepackage{subfigure}
\usepackage{soul}
\usepackage{physics}

\newcommand{\stkout}[1]{\ifmmode\text{\sout{\ensuremath{#1}}}\else\sout{#1}\fi}
\newcommand{\e}{{\rm e}}
\newcommand{\imai}{{\rm i}}

\usepackage{color}

\newcommand{\be}{\begin{equation}}
\newcommand{\bw}{\begin{widetext}}
\newcommand{\ew}{\end{widetext}}
\newcommand{\bea}{\begin{eqnarray}}
\newcommand{\bc}{\begin{center}}            
\newcommand{\ee}{\end{equation}}
\newcommand{\eea}{\end{eqnarray}}
\newcommand{\ec}{\end{center}}

\begin{document}


\title{Violation of kinetic uncertainty relation in  maser heat engines: Role of spontaneous emission}

\author{Varinder Singh}
\affiliation{School of Physics, Korea Institute for Advanced Study, Seoul 02455, Republic of  Korea}

\author{Euijoon Kwon}
\affiliation{School of Physics, Korea Institute for Advanced Study, Seoul 02455, Republic of  Korea}
\affiliation{Department of Physics and Astronomy \& Center for Theoretical Physics, Seoul National University, Seoul 08826, Republic of Korea} 

\author{Jae Sung Lee}
\affiliation{School of Physics, Korea Institute for Advanced Study, Seoul 02455, Republic of  Korea}

\date{\today}

\begin{abstract}
We investigate the kinetic uncertainty relation (KUR)—a fundamental trade-off between dynamical activity and current fluctuations—in two configurations of a maser heat engine. We find that KUR violations arise only in one model. This asymmetry originates from spontaneous emission, which breaks the structural symmetry between the configurations and modifies their coherence dynamics. While we analyze several contributing factors—including statistical signatures such as the Fano factor and the ratio of dynamical activity to current—our results show that the decisive mechanism is the slower decoherence in one configuration, which enables quantum violations of the classical steady-state KUR bound. By contrast, the faster coherence decay in the other configuration suppresses such violations, driving it closer to classical behavior. These findings highlight the critical role of decoherence mechanisms in determining fundamental thermodynamic bounds and provide insights for the design of quantum heat engines in which the control of decoherence is central to suppressing fluctuations and enhancing reliable performance.

\end{abstract}

\maketitle
\section{Introduction}
Since the industrial revolution, heat engines have served as a cornerstone in the development of classical thermodynamics, shaping both theoretical insights and experimental innovations. Today, as we extend thermodynamics into the quantum realm, quantum heat engines continue to occupy a central place—especially within the rapidly growing field of quantum thermodynamics \cite{Kosloff2013,Sai2016,Sourav2021,Mahler,AlickiKosloff,DeffnerBook}. These quantum thermal machines are not only crucial for deepening our understanding of energy conversion at the nanoscale but also for their potential role as building blocks of future quantum technologies.

In this context, the ability to harness uniquely quantum features—such as coherence, entanglement, and engineered quantum reservoirs—offers new pathways for enhancing the performance of heat engines beyond classical limits. However, thermal machines at the nanoscale are inherently subject to significant fluctuations, both thermal and quantum, which can degrade stability and reduce efficiency. Suppressing these fluctuations, or at least understanding their fundamental constraints, is therefore vital for realizing reliable and efficient quantum technologies.

A key step toward this goal was the introduction of the thermodynamic uncertainty relation (TUR) \cite{Barato2015,Gingrich2016,Horowitz2017,Dechant2018,Hasegawa2019,Koyuk2019,Koyuk2020,Buffoni2020,Horowitz2020,Lee2021,Shiraishi2021,Vo2022,Razzoli2024,Kwon2024,Chesi2025,BrandnerSaito2025}, which quantifies a universal trade-off between the precision of output currents (e.g., power) and the entropy production rate required to maintain nonequilibrium operation. Originally formulated for classical steady-state systems, the TUR sets a lower bound on current fluctuations in terms of thermodynamic cost. In the context of steady-state heat engines, the TUR can be written as
\begin{equation}
 \frac{\langle\sigma\rangle}{k_{\rm B}}\frac{\Delta P}{\langle P\rangle^ 2}\ge 2, \label{TUR}
\end{equation}
where  $\langle P \rangle$  and $\Delta P=\text{lim}_{t\rightarrow\infty} \langle [P(t) -  \langle P \rangle ]^2\rangle t$
denote the mean power and rescaled variance of the power in the steady state and $k_{\rm B}$ is the Boltzmann constant.  

Another class of uncertainty relations, complementary to the TUR, known as kinetic uncertainty relations (KURs) \cite{Garrahan2017,DiTerlizzi2019}, places bounds on current fluctuations based on a system’s dynamical activity rather than its thermodynamic cost. The dynamical activity, denoted by $A$, quantifies the average number of transitions or microscopic events occurring per unit time in a stochastic system. In this work, we focus on studying the KUR within the context of a heat engine setup. Originally, the KUR was derived for time-homogeneous Markov jump processes and is expressed as follows \cite{Garrahan2017,DiTerlizzi2019}
\begin{equation}
 A\frac{\Delta P}{\langle P\rangle^ 2}\ge 1. \label{KUR}
\end{equation}

In the quantum domain, however, the TUR and the KUR can be violated. While much attention has been given to violations of the TUR~\cite{Patrick2021,Menczel2021,Davinder2021}, the exploration of KUR violations in quantum systems began only recently \cite{Prech2025,VuSaito2022,Hasegawa2020,Vu2025PRX,EuijoonJS2024,Prech2023,MoreiraMark2025,Palmqvist2025}.  It is important to note that quantum effects do not always reduce fluctuations; in some cases, they can actually amplify them \cite{Mark2021}. Therefore, it is prudent to examine  the mechanisms underlying the violations of the KUR.
  \begin{figure*}    
 \begin{center}
 \includegraphics[width=1\textwidth]{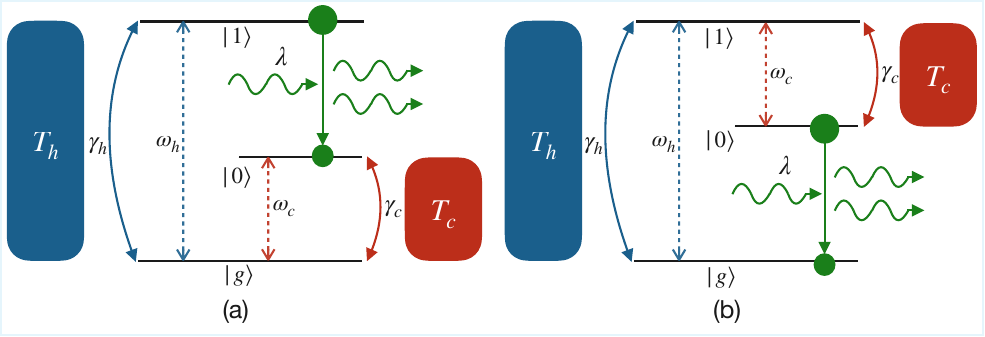}
 \end{center}
\caption{(a) Model I describes a three-level maser heat engine that operates continuously while interacting with two thermal reservoirs at temperatures $T_h$  (hot) and $T_c$  (cold), with corresponding coupling strengths $\gamma_h$ and $\gamma_c$. The system also interacts with a classical single-mode field, with the interaction strength characterized by the coupling constant $\lambda$, which governs the matter-field interaction. (b) Model II presents a variation of Model I. In this configuration, the cold reservoir is coupled to the two upper energy levels instead of the two lower ones as in Model I. Additionally, the mechanism responsible for power extraction is now coupled to the two lower levels rather than the upper levels, marking a key structural difference from the first model. }
\end{figure*} 

 In this work, we explore the validity of the KUR in Eq. (\ref{KUR}) in the three-level maser heat engine \cite{Scovil1959,VJ2019,Varinder2020,Geva1994,Geva1996,BoukobzaTannor2007,BoukobzaTannor2006A,VJ2020,Kiran2021}, a experimentally realizable quantum heat engine   \cite{Klatzow2019}. This model was the first quantum heat engine proposed and is commonly referred to as the SSD engine, named after the authors Scovil and Schulz-DuBois \cite{Scovil1959}. Given its significance both experimentally and theoretically, there is considerable interest in reducing fluctuations in the power output of the SSD engine \cite{Patrick2021,VSTUR2023,BrijMohan2025}. 
 
 This motivates a thorough investigation of the model in the context of KURs. To this end, we will first demonstrate that the violation of the KUR is highly sensitive to the spontaneous emission phenomenon. Specifically, we will investigate two slightly different variants of the three-level heat engine, distinguished by the levels that couple to the thermal reservoirs and the levels that interact with the electromagnetic field. We will show that even this small difference leads to significantly different outcomes when evaluating the KUR in both SSD variants.

The paper is organized as follows. In Section \ref{sec:model_def}, we provide an overview of the SSD model. Section \ref{sec:TUR_3_level} explores the violations of KUR in two slightly different variants of the SSD model, comparing the extent of the violations and investigating the physical origin of why the violation occurs in only one of them. We conclude in Section \ref{sec:conclusions}.

\section{The SSD model}
\label{sec:model_def}

The SSD engine~\cite{Scovil1959} serves as a conventional and widely studied example of a quantum heat engine. It consists of a three-level quantum system simultaneously interacting with two thermal reservoirs at different temperatures, $T_c$ and $T_h$, where 
$T_c<T_h$. In the first configuration, referred to as Model I (see Fig. 1(a)), the hot reservoir drives the excitation between the ground state $\ket{g}$ and excited state  $\ket{1}$, while the cold reservoir facilitates relaxation between the states $\ket{0}$ and $\ket{g}$.  The interaction between the system and an electromagnetic field of frequency $\omega$ is described by the following semiclassical Hamiltonian, formulated within the rotating wave approximation:
$V(t)=  \lambda (e^{-i\omega t} \vert 1\rangle\langle 0\vert + e^{i\omega t} \vert 0\rangle\langle 1\vert)$; 
Here, $\lambda$ denotes the field-matter coupling strength. In our study, we focus on the resonant case where the single-mode field matches the energy difference between the lasing states $\ket{0}$ and $\ket{1}$, i.e., $\omega=\omega_1-\omega_0$.

In Model II of the SSD engine (see Fig. 1(b)), the cold reservoir is coupled to the transition between the states  $\ket{0}$  and $\ket{1}$, while the electromagnetic field interacts with the $\ket{g} \leftrightarrow \ket{0}$ transition. Consequently, the driving frequency is  $\omega=\omega_0-\omega_g$. All other aspects of the setup remain identical to those in Model I. Although this modification appears minor, it leads to several interesting consequences, which will be explored in the subsequent sections.

The dynamics of the three-level system, viewed in a frame rotating with the system Hamiltonian $H_0$, is captured by the Lindblad master equation given below:
\begin{equation}
\dot{\rho} = - i [V_R,\rho] + \mathcal{L}_{h}[\rho] + \mathcal{L}_{c}[\rho]. \label{LindbladMain}
\end{equation}
In Model I, the interaction Hamiltonian is given by $V_R^{\rm I}= \lambda(\vert 1\rangle\langle 0\vert + \vert 0\rangle\langle 1\vert)$,  while for Model II, it takes the form $V_R^{\rm II}= \lambda(\vert g\rangle\langle 0\vert + \vert 0\rangle\langle g\vert)$.  The term $\mathcal{L}_{h,c}$
 represents the interaction of the system with the hot (cold) thermal reservoir:
\begin{eqnarray}
\mathcal{L}_h[\rho ] &=& \gamma_h(n_h+1) \Big (\sigma_{g1} \rho \sigma^{\dagger}_{g1} -\frac{1}{2}  \{\sigma^{\dagger}_{g1} \sigma_{g1},\rho\}   \Big) \nonumber
\\
&& +\gamma_h n_h  \Big(\sigma^{\dagger}_{g1} \rho \sigma_{g1} -\frac{1}{2}  \{\sigma_{g1} \sigma^{\dagger}_{g1} ,\rho\}   \Big) ,
\label{D1}
\end{eqnarray} 
\begin{eqnarray}
\mathcal{L}_c[\rho] &=& \gamma_c(n_c+1) \Big (\sigma_{ab} \rho \sigma^{\dagger}_{ab} -\frac{1}{2}  \{\sigma^{\dagger}_{ab} \sigma_{ab},\rho\}   \Big) \nonumber
\\
&& +\gamma_c n_c  \Big (\sigma^{\dagger}_{ab} \rho \sigma_{ab} -\frac{1}{2}  \{\sigma_{ab} \sigma^{\dagger}_{ab} ,\rho\}   \Big) , \label{D2}
\end{eqnarray}
where $\sigma_{ab}=\ket{a}\bra{b}$, and   $\gamma_c$,  $\gamma_h$ denote the coupling strengths between the system and the cold and hot reservoirs, respectively. The hot reservoir dissipator $\mathcal{L}_h[\rho]$ is the same for both models, as it acts on the same transition $\ket{g}\leftrightarrow\ket{1}$. However, the cold reservoir dissipator $\mathcal{L}_c[\rho]$  differs between the two models due to the distinct transitions involved. In Model I, it acts on the $\ket{g}\leftrightarrow\ket{0}$ transition, so $\sigma_{ab}= \sigma_{g0}$. In contrast, in Model II, the cold reservoir couples the levels $\ket{0}\leftrightarrow\ket{1}$, and thus $\sigma_{ab}= \sigma_{01}$.  
 Finally, $n_{h}= 1/(\exp[ \omega_{h}/  T_{h}]-1)$ and  $n_{c}= 1/(\exp[ \omega_{c}/  T_{c}]-1)$ denote the average photon occupation numbers at frequencies $\omega_h$ and $\omega_c$  in the hot and cold baths, respectively.
 For Model I (Model II), the cold transition frequency is given by  $\omega_c=\omega_0-\omega_g$ $\left(\omega_c=\omega_1-\omega_0\right)$, while the hot transition frequency is the same in both models, $\omega_h=\omega_1- \omega_g$.

\begin{figure}   
 \begin{center}
\includegraphics[width=8.6cm]{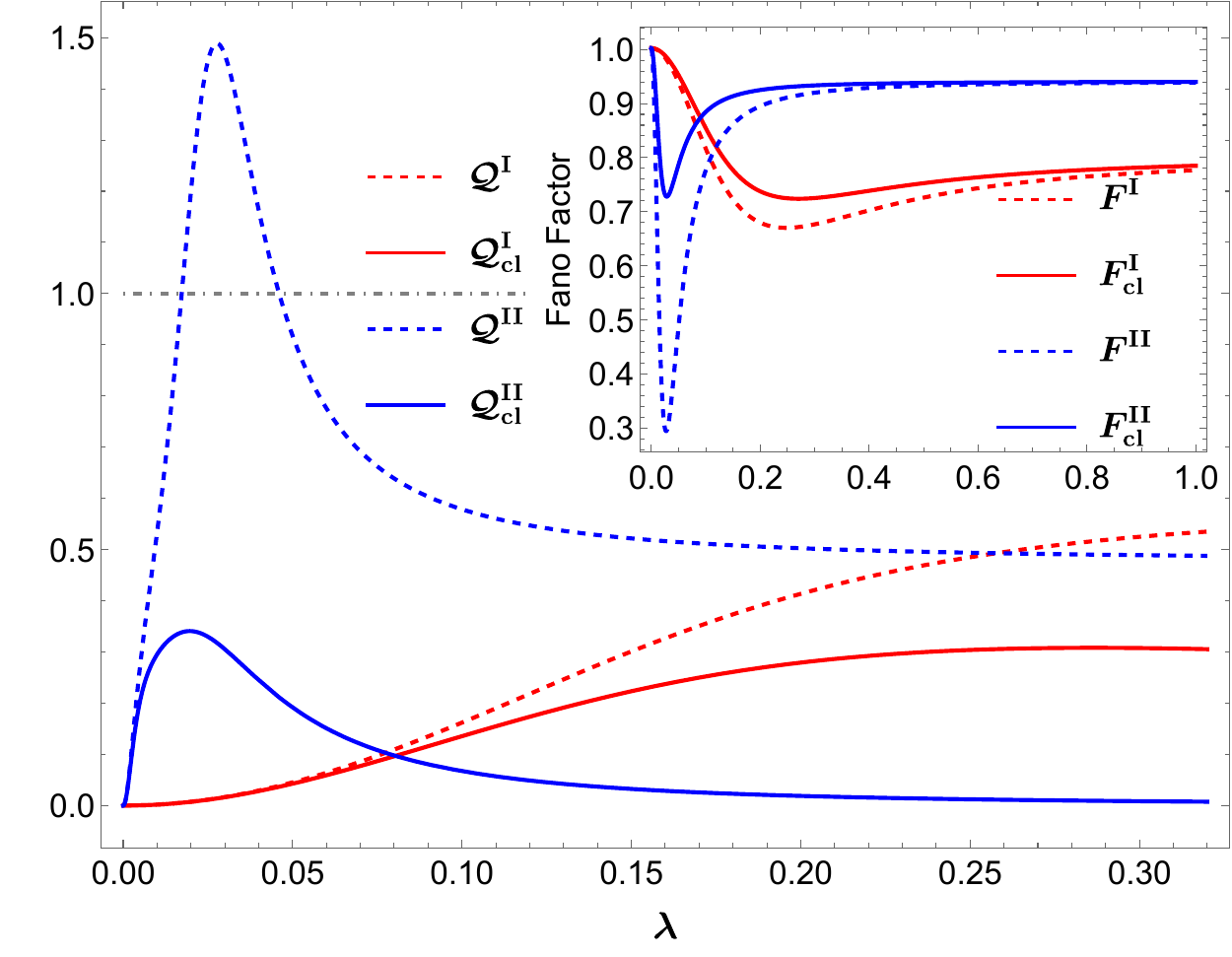}
 \end{center}
\caption{$\mathcal{Q}$ as a function of the matter–field coupling $\lambda$. Red and blue curves correspond to Model I and Model II, respectively, with parameters fixed at $\gamma_h=0.016$, $\gamma_c=2$, $n_h=5$, and $n_c=0.001$. In the main panel, dotted curves (red/blue) show Eqs. (\ref{KURme}) and (\ref{KURPatrick}), while solid curves show the corresponding classical equivalents. The inset displays the Fano factors from Eqs.(\ref{FanoI}) and (\ref{FanoII}) (dotted), alongside their classical equivalents (solid).} \label{KURgraph}
\end{figure}
 
%

\section{  KUR violations in the SSD model}
\label{sec:TUR_3_level}
In this section, we examine the SSD model through the lens of the KUR and compare two closely related variants of the model. The difference between these variants lies in the specific energy levels connected by the cold reservoir in the three-level system: in the first variant, the cold reservoir couples the states  $\ket{g}$ and $\ket{0}$, whereas in the second, it links $\ket{0}$ and $\ket{1}$.
Despite this distinction, both configurations are commonly treated as equivalent in the literature and are generally referred to by the same name—the SSD model.
Although these two configurations appear similar, our analysis of the KUR reveals that they lead to fundamentally different results. This discrepancy arises entirely from the quantum effect of spontaneous emission, which plays distinct roles in each configuration.

\subsection{KUR Violations}
In Model I, for example, the system absorbs a photon from the hot reservoir, performs work via stimulated emission, and then emits a photon with the remaining energy into the cold reservoir. The net current $I$ in the system is defined as the average number of such thermodynamic cycles proceeding in the nominal direction ($\ket{g}\rightarrow \ket{1}\rightarrow \ket{0}\rightarrow \ket{g}$) per unit time, minus the number of cycles occurring in the reverse direction. The net current can be defined in the same way for Model II. 

The power output of the engine is obtained by multiplying the energy of the emitted photon used as work with the net current of such transitions, and is given by:
\begin{equation}
    P = (\omega_h-\omega_c) \langle I\rangle. \label{aux1}
\end{equation}
Similarly, the  variance in power is obtained as follows~\cite{VSTUR2023,Patrick2021}:
 \begin{equation}
     \Delta P = (\omega_h-\omega_c)^2  \Delta I. \label{aux2}
 \end{equation}
Using Eqs. (\ref{aux1}) and  (\ref{aux2}), KUR given in Eq. (\ref{KUR}) can be written in the form
\begin{equation}
    A \frac{\Delta I}{ \langle I\rangle^2} \geq 1. \label{KUR2}
\end{equation}
However, for our analysis, it is more suitable to cast the above inequality in the following form:
\begin{equation}
    \mathcal{Q} \equiv \frac{ \langle I\rangle^2 }{A\,\Delta I} \leq 1, \label{KURinverse}
\end{equation}
where we introduced $\mathcal{Q}$ as KUR quantifier. All the quantities appearing in Eq. (\ref{KURinverse})   can be systematically obtained  in the long-time limit using the full counting statistics (FCS) formalism outlined in Appendix B. For Model I,  FCS yields the following analytical expression for the KUR quantifier:
%
%
\begin{widetext}
\begin{equation}
      \mathcal{Q}^{\rm I}  =   \frac{2(n_h-n_c)^2\gamma_c\gamma_h\lambda^2}{B\,B'\,H }
\bigg[ 
        2n_h n_c +n_h+n_c - \frac{8  \lambda^2  (n_h-n_c)^2 \gamma_c\gamma_h G}{(B\,C\, \gamma_c\gamma_h  \,+\, 4\lambda^2 D)^2 }
\bigg]^{-1}, \label{KURme}
\end{equation}
where 
$B = \gamma_c(1+n_c) + \gamma_h(1+n_h)$,   $B' = \gamma_c n_c + \gamma_h n_h$, 
$C =3 n_c n_h+2 n_c+2 n_h+1$,
$D= \gamma_c(1+3n_c) + \gamma_h(1+3n_h)$,
$G =8 \lambda ^2+ \gamma _c \gamma _h \left(  10 n_h n_c +7n_c +7 n_h+4\right)+\gamma _c^2 \left(n_c+1\right) \left(2 n_c+1\right)+\gamma _h^2 \left(n_h+1\right) \left(2 n_h+1\right)$,
$H=4\lambda^2 + (1+n_c)(1+n_h)\gamma_c\gamma_h$.
\vspace{2mm}

Similarly, for Model II, the KUR   quantifier $\mathcal{Q}^{\rm II}$ is given by
\begin{equation}
\mathcal{Q}^{\rm II} =   \frac{2(n_h-n_c)^2\gamma_c\gamma_h\lambda^2}{B\,B'\,H' }  \Bigg[2n_h n_c +n_h+n_c - \frac{8\lambda^2(n_h-n_c)^2 \gamma_c\gamma_h G'}{(B'\,C'\, \gamma_c\gamma_h  \,+\, 4\lambda^2 D')^2 }
\Bigg]^{-1},\label{KURPatrick}
\end{equation}
where $C' =3 n_c n_h+ n_c+ n_h$, $D'= \gamma_c(2+3n_c) + \gamma_h(2+3n_h)$, $G'=8 \lambda ^2+\gamma _c \gamma _h \left(10 n_h n_c+3 n_c+3 n_h\right)+\gamma _c^2 n_c \left(2 n_c+1\right)+\gamma _h^2 n_h \left(2 n_h+1\right)$, $H'=4\lambda^2 + n_c n_h\gamma_c\gamma_h$.
\end{widetext} 
	\begin{figure*}
		\centering
		\includegraphics[width=.95\textwidth]{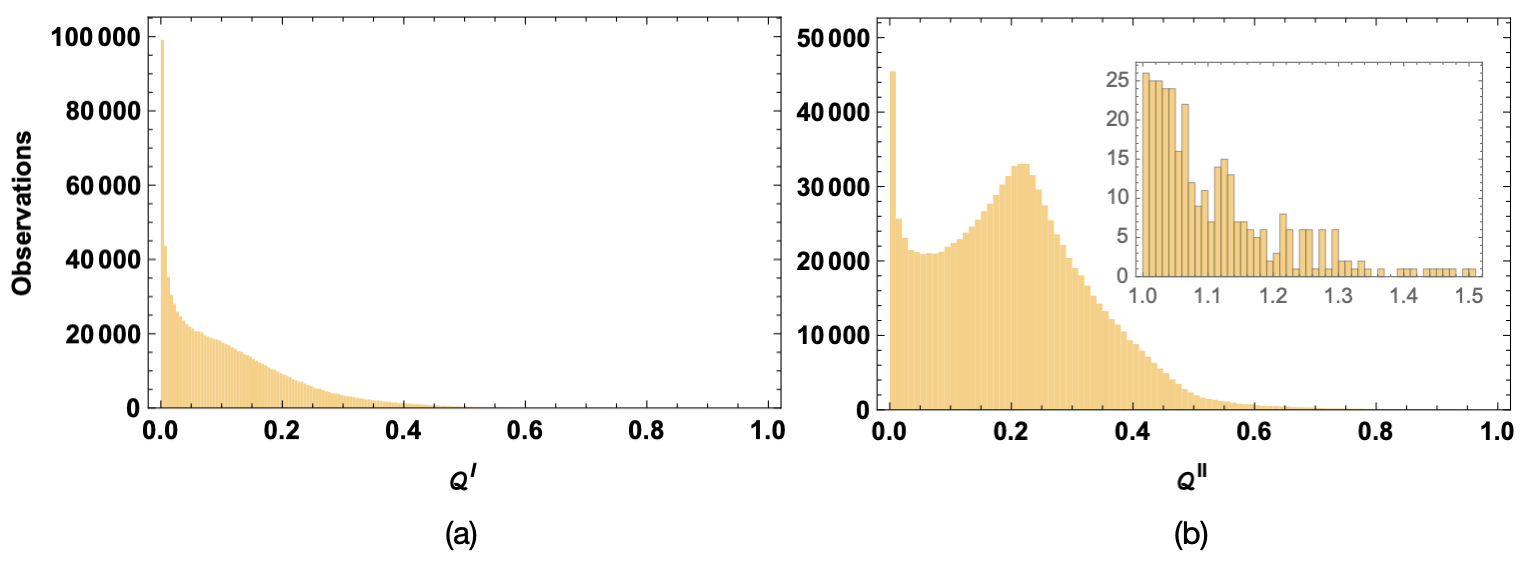}
		\caption{Histograms of sampled values of $\mathcal{Q}^{\rm I}$ (Eq. (\ref{KURme})) and $\mathcal{Q}^{\rm II}$   (Eq. (\ref{KURPatrick})) for random sampling over a region of the parametric space.  The insets show the subset of the sampled data for which KUR violations are happening. The parameters are sampled over the uniform distributions $ \gamma_{h}\in [10^{-4},1]$, $ \gamma_{c}\in [10^{-4},5]$, $n_{h}\in [0,5]$, $n_{c}\in [0,.1]$ and $\lambda\in[10^{-4},2]$. For plotting the histograms, we choose a bin width of 0.01 to arrange $10^6$ data points.} \label{hist}
	\end{figure*}
We plot Eqs. (\ref{KURme}) and (\ref{KURPatrick}) as functions of the matter-field coupling $\lambda$, keeping all other parameters fixed. Interestingly, violations of the KUR are observed only in Model II (dotted blue curve), whereas in Model I (dotted red curve), the KUR remains satisfied across the entire range of $\lambda$. For larger values of $\lambda$, both $\mathcal{Q}^{\rm I}$ and $\mathcal{Q}^{\rm II}$ saturate to values well below the KUR bound, thereby satisfying the inequality. We have verified this behavior over a wide range of parameter values and consistently found no violations of the KUR in Model I.

To further investigate KUR violations across a broad parameter space, we randomly sample values within a specified region and generate histograms of $\mathcal{Q}^{\rm I}$ and $\mathcal{Q}^{\rm II}$, as shown in Fig.~\ref{hist}. 

The violation of the  steady-state classical KUR occurs exclusively in Model II (see Inset of Fig. 3), whereas all sampled values of $\mathcal{Q}^{\rm I}$ for Model I  lie well below the KUR bound.


\subsection{Factors leading to KUR violations only in Model II}
To analyze why KUR violations arise only in Model II, it is convenient to rewrite Eq. (\ref{KUR2}) in the   form
\begin{equation}
    \frac{A}{\langle I \rangle} F \geq 1, \label{RFano}
\end{equation}
where $F\equiv\Delta I/\langle I\rangle$ is the Fano factor. Since $I$ counts directional jumps and $A$ counts total number of jumps, we have $A/\langle I\rangle\geq 1$, thus the classical KUR  can be violated in quantum systems only when the Fano factor $F \leq 1$. This condition is necessary, though not sufficient, for such a violation to occur.  

We now examine the two key factors appearing in Eq.~(\ref{RFano}): the ratio of dynamical activity to current, denoted as 
$R\equiv A/\langle I \rangle$, and the Fano factor $F$. Lower values of both $R$ and $F$ increase the likelihood of violating the KUR in the maser heat engine setup. Our analysis shows that, for fixed values of all other parameters, a smaller Fano factor favors lower values of $n_c$.  In particular, when $n_c=0$,  Fano factors for Model I and Model II can be obtained by substituting $n_c=0$ in Eqs. (\ref{FanoI}) and (\ref{FanoII}), as derived in Appendix B:
\begin{widetext}
    \begin{equation}
        F^{\rm I} = 1-\frac{8 \lambda ^2 \gamma _c^2 n_h \left(\gamma _c^2+\gamma _c^2 \left(n_h+1\right) \left(2 n_h+1\right)+\gamma _c^2 \left(7 n_h+4\right)+8 \lambda ^2\right)}{\left(4 \lambda ^2 \left(2 \gamma _c+3 \gamma _c n_h\right)+\gamma _c^2 \left(2 n_h+1\right) \left(2 \gamma _c+\gamma _c n_h\right)\right){}^2} \leq 1, \label{Fano1}
    \end{equation} 
    \begin{equation}
        F^{\rm II} = 1 - \frac{8 \lambda ^2 \gamma _c \gamma _h n_h \left(\gamma _h n_h \left(3 \gamma _c+\gamma _h \left(2 n_h+1\right)\right)+8 \lambda ^2\right)}{\left(\gamma _c \left(\gamma _h^2 n_h^2+8 \lambda ^2\right)+4 \lambda ^2 \gamma _h \left(3 n_h+2\right)\right){}^2} \leq 1.  \label{Fano2}
    \end{equation}
We clearly observe that both $F^{\rm I}$ and $F^{\rm II}$ remain strictly below 1 when $n_c = 0$, indicating that this is the most favorable condition for KUR violations in maser heat engines as Fano factor greater than 1 never yield KUR violations as discussed earlier. Consequently, we choose numerical values of $n_c$ close to zero in order to systematically explore the regime where KUR violations are most prominent in our setup.  For our analysis, we plot Eqs.~(\ref{FanoI}) and (\ref{FanoII})  in the inset of Fig.~2, shown as dotted blue and dotted red curves, respectively, using the same parameters as in the main panel of Fig.~2. It is evident that for smaller values of $\lambda$
(which favor KUR violations), the Fano factor for Model II is significantly lower than that of Model I, making Model II more susceptible to KUR violations.
	
Let us now analyze the behavior of the ratio of dynamical activity to current in both models. The analytical expressions for the corresponding ratios are derived to be
\begin{equation}
    R^{\rm I} =\frac{A^{\rm I}}{\langle I^{\rm I} \rangle} = \frac{\left(\gamma _c n_c+\gamma _h n_h\right) \left(\gamma _c \left(n_c+1\right)+\gamma _h \left(n_h+1\right)\right) \left(\gamma _c \left(n_c+1\right) \gamma _h \left(n_h+1\right)+4 \lambda ^2\right)}{2 \lambda ^2 \gamma _c \gamma _h \left(n_h-n_c\right)} , \label{R1}
\end{equation}
\begin{equation}
      R^{\rm II} = \frac{A^{\rm II}}{\langle I^{\rm II} \rangle} = \frac{\left(\gamma _c n_c+\gamma _h n_h\right) \left(\gamma _c \left(n_c+1\right)+\gamma _h \left(n_h+1\right)\right) \left(\gamma _c n_c \gamma _h n_h+4 \lambda ^2\right)}{2 \lambda ^2 \gamma _c \gamma _h \left(n_h-n_c\right)}. \label{R2}
\end{equation} 
\end{widetext}
Comparing Eqs. (\ref{R1}) and  (\ref{R2}), we find that $R^{\rm I}\geq R^{\rm II}$. Thus, we conclude that both the Fano factor and the ratio of dynamical activity to current attain lower values in Model II, thereby making it more favorable for KUR violations.
\begin{figure}
		\centering
		\includegraphics[width=.45\textwidth]{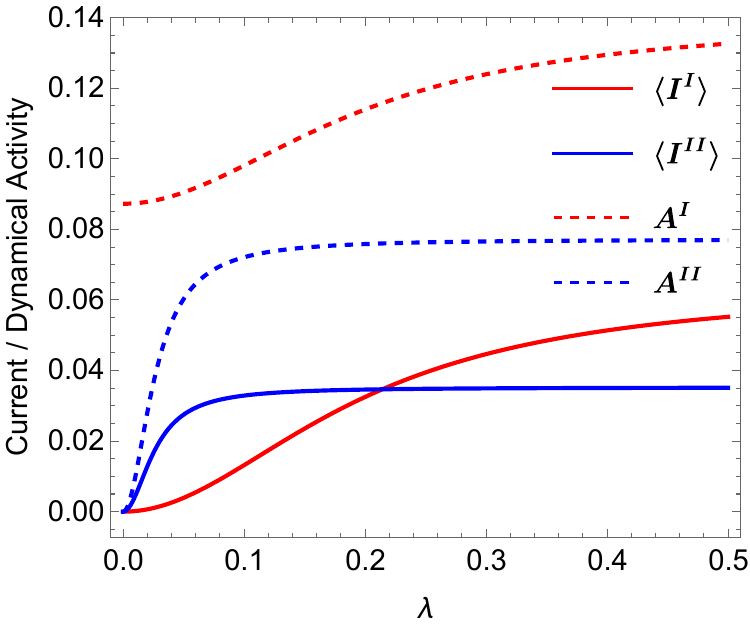}
		\caption {Current and dynamical activity as a function of matter-field coupling parameter $\lambda$ for both Model I and Model II. The solid red and blue curves represent the current in Model I and Model II, respectively, while the corresponding dotted curves depict the dynamical activity in each model. The parameters used are same as in Fig. 2.  } \label{CurrentDAFano}
	\end{figure}

Finally, we examine the behavior of the dynamical activity and the current separately (see Fig. 4).  
As the field is switched on (i.e., as $\lambda$ increases from zero to finite values), we observe that the current in Model II grows more rapidly than in Model I, eventually saturating to a fixed value in both cases.   Furthermore, in Model II, the dynamical activity increases from zero to a finite value, whereas in Model I, it grows from one finite value to another. Since KUR violations in Fig. 2 occur predominantly at lower values of $\lambda$ ($\lambda<0.1$) for $n_c=0.001$, our focus will be on understanding the physics within this regime. This can be achieved by analyzing the limiting behavior of the steady-state populations in both models in the limit $n_c = 0$ and $\lambda = 0$. The expressions for the steady-state populations in Model I and Model II, evaluated in the limit $n_c = 0$ and $\lambda = 0$, are given by the following equations, respectively:
\begin{equation}
    \rho_{gg}^{\rm I} = \frac{1+n_h}{1+2n_h}, \quad \rho_{00}^{\rm I}  = 0, \quad \rho_{11}^{\rm I}  = \frac{n_h}{1+2n_h},
\end{equation}
and
\begin{equation}
    \rho_{gg}^{\rm II} = 0, \quad \rho_{00}^{\rm II}  = 1, \quad \rho_{11}^{\rm II}  = 0.
\end{equation}
In Model I, in the absence of the field, levels $\ket{0}$ and $\ket{1}$ are decoupled. Thus, transitions coupling levels $\ket{g}$ and $\ket{1}$ ($\ket{g}$ and $\ket{0}$) will be at thermal equilibrium with their respective baths. Since $n_c = 0$ is chosen, the two-level subsystem with states $\ket{g}$ and $\ket{0}$ will have all its population settled in the ground state $\ket{g}$ due to the absence of upward transitions. This explains the vanishing population in state $\ket{0}$, i.e., $\rho^{\rm I}_{00} = 0$. 
The finite value of dynamical activity—despite a vanishing current—is solely due to the jumps occurring within the two-level subsystem formed by the states $\ket{g}$ and $\ket{1}$, which helps maintain overall thermal equilibrium in the system. 

In contrast to Model I, where the levels $\ket{0}$ and $\ket{1}$ are decoupled in the absence of the field, Model II features a decoupling between $\ket{g}$ and $\ket{0}$. This structural difference leads to interesting consequences for the engine's performance, particularly in the context of KUR violations. In Model II, in the absence of the driving field, the steady-state population accumulates in the state  $\ket{0}$. The system undergoes an upward transition from $\ket{g}$ to $\ket{1}$, followed by a downward transition from $\ket{1}$ to $\ket{0}$, the latter occurring purely via spontaneous emission. Once the entire population settles into    $\ket{0}$, all transitions cease, and the system becomes dynamically inactive resulting in vanishing dynamical activity. The same reasoning helps explain the faster growth of current in Model II. Owing to the specific way the reservoirs are coupled to different energy levels in the two models, complete population inversion between lasing levels can be achieved only in Model II. This enhanced inversion directly contributes to the more rapid increase in current observed in Model II compared to Model I.

\subsection{Equivalent Classical Models}
A classical equivalent of both Model I and Model II can be constructed by  appropriately modifying the coherent dynamics~\cite{Sprekeler2004,Kiesslich2006,Patrick2021,Prech2025}. We begin with Model I, where the classical analogue is obtained by replacing the coherent driving term $-\frac{i}{\hbar} [V_R^{\rm I},\rho]$ in Eq.~(\ref{LindbladMain}) with an incoherent counterpart. Specifically, the replacement is given by  
\begin{equation}
    -\frac{i}{\hbar} \left[V_R^{\rm I},\rho\right] \rightarrow \gamma_{\rm cl}^{\rm I} \left( \mathcal{D}_{\sigma_{10}}[\rho_{\rm cl}]  +  \mathcal{D}_{\sigma_{01}} [\rho_{\rm cl}]\right) , \label{V_R_replacement}
\end{equation} 
where 
$\gamma_{\rm cl}^{\rm I} $ is classical rate constant,  and the dissipator is defined as
$\mathcal{D}_{\sigma}[\rho_{\rm cl}]=\sigma\rho_{\rm cl}\sigma^\dagger-\frac{1}{2}\{\sigma^\dagger \sigma,\rho_{\rm cl}\}$.  
For Model I, the resulting  master equation with the replacement Eq.~\eqref{V_R_replacement} can be tuned to reproduce the same current and populations as the quantum model [Eqs.~(\ref{CurrentI}) and (\ref{rhoggI})–(\ref{rho11I})]  by setting $\gamma^{\rm I}_{\rm cl} = 4\lambda^2/(\gamma_h(n_h+1)+\gamma_c (n_c+1))$  (see Appendix C for details). However, the dynamical activity and Fano factor differ between the quantum model and its classical counterpart.

Similarly, for Model II, replacing the coherent Hamiltonian term with $\gamma^{\rm II}_{\rm cl} ( \mathcal{D}_{\sigma_{g0}}  +  \mathcal{D}_{\sigma_{0g}})$, yields a classical master equation whose steady-state current and populations coincide with those of the quantum model [Eqs.~(\ref{CurrentII}), (\ref{rhoggII})–(\ref{rho11II})], when the classical rate is chosen as
  $\gamma^{\rm II}_{\rm cl} = 4\lambda^2/(\gamma_h n_h +\gamma_c  n_c)$. 

 Using the full counting statistics, we obtain explicit expressions for the Fano factors of the classical models under consideration. We can analytically show that the corresponding classical models always yield larger Fano factors compared to their quantum counterparts. The difference  $\Delta F^{k} = F^{\rm k}_{\rm cl} - F^{\rm k}  \,(k=\rm I,\,\rm II)$  between  the classical and quantum Fano factors is given by
\begin{widetext}
\begin{eqnarray}
  \Delta F^{\rm I}&=&  \frac{16 \lambda ^2 \gamma _c^2 \gamma _h^2 \left(n_h-n_c\right) \left(  3 n_h n_c+2 n_h+2n_c+1\right)}{\left(4 \lambda ^2 \left(\gamma _c+3 \gamma _c n_c+\gamma _h+3 \gamma _h n_h\right)+\gamma _c \gamma _h \left(n_c \left(3 n_h+2\right)+2 n_h+1\right) \left(\gamma _c+\gamma _c n_c+\gamma _h+\gamma _h n_h\right)\right){}^2}  \nonumber
  \\
  &=&
  \frac{   3 n_h n_c+2 n_h+2n_c+1}{\lambda^2(n_h-n_c)} \langle I_{\rm I}\rangle^2 ,
\end{eqnarray}
and 
\begin{eqnarray}
  \Delta F^{\rm II} &=&   \frac{16 \lambda ^2 \gamma _c^2 \gamma _h^2 \left(n_h-n_c\right) \left(3 n_c n_h+n_c+n_h\right)}{\left(4 \lambda ^2 \left(\gamma _c \left(3 n_c+2\right)+\gamma _h \left(3 n_h+2\right)\right)+\gamma _c \gamma _h \left(3 n_c n_h+n_c+n_h\right) \left(\gamma _c n_c+\gamma _h n_h\right)\right){}^2} \ \nonumber \\
  &= &
  \frac{3n_h n_c + n_c+n_h}{\lambda^2(n_h-n_c)} \langle I_{\rm II}\rangle^2 \,,   \label{DeltaFano2}
\end{eqnarray}
\end{widetext}
respectively. For $n_h>n_c$, which is always true for three-level maser setup operating as a heat engine, 
the above equations imply that  $F^{\rm I, \,II} \leq F^{\rm I, \,II}_{\rm cl}$.   Similarly, expressions for the  dynamical activities for equivalent classical models can be found. Once all relevant quantities for the classical models are obtained, the corresponding KUR relations can be derived. However, their analytical expressions are lengthy and not particularly insightful, so we do not present them here. Instead, we display the KUR values for both classical models in Fig. 2 (solid red and blue curves). It is evident that both classical analogs not only satisfy the classical KUR bound, but their values consistently lie below those of their quantum counterparts. Given the dependence of the classical KUR expressions on many parameters, an analytical proof is intractable. Nevertheless, we have also examined corresponding histograms (not shown here to avoid redundancy with Fig. 3), and in none of the sampled events do the classical models exhibit any KUR violations.

An analysis of the Fano factors in the inset of Fig. 2 reveals that the Fano factor for Model I remains closer to that of its classical counterpart, whereas Model II shows a more pronounced deviation. This deviation, along with the observation that KUR violations occur only in Model II, suggests that Model II exhibits a higher degree of quantum behavior compared to Model I. This can be understood by looking at the decoherence rates in both models. We have
   \begin{equation}
      \dot{\rho}^{\rm I}_{10} =  -\frac{1}{2}[\gamma_h(n_h+1)+\gamma_c(n_c+1)]\rho_{10} + i\lambda(\rho_{11}-\rho_{00}), \label{decoherence1}
    \end{equation}
       \begin{equation}
      \dot{\rho}^{\rm II}_{g0} = - \frac{1}{2}(\gamma_h n_h + \gamma_c n_c ) \rho_{g0} + i\lambda(\rho_{00}-\rho_{gg}). \label{decoherence2}
    \end{equation}
It is evident that Model I exhibits a faster decay of coherence compared to Model II. This is primarily because its decoherence rate includes additional contributions from both $\gamma_h$ and $\gamma_c$, which originate from spontaneous emission and zero-point fluctuations. These extra channels of environmental noise accelerate the loss of quantum coherence in Model I. As a result, Model I behaves more classically, maintaining a closer resemblance to its classical counterpart than Model II, which retains a higher degree of quantumness due to its relatively slower decoherence decay.

\section{Conclusions} \label{sec:conclusions}
In this work, we investigated the KUR in two closely related configurations of a three-level maser heat engine. Despite their structural similarity, we found that KUR violations occur exclusively in one of the two models. This asymmetry originates from the different roles played by spontaneous emission in each configuration, which significantly affects the coherence dynamics and, consequently, the fluctuation behavior of the engine.

Our analysis shows that Model II, which exhibits slower decoherence due to the absence of additional spontaneous emission terms in the decoherence rate, shows KUR violations under certain conditions—particularly for low values of the matter-field coupling and when the cold bath occupation number $n_c$ approaches zero. In contrast, Model I, with a faster coherence decay and stronger influence of environmental noise, remains consistently within the bounds of classical KUR behavior.

By comparing the Fano factors and the ratio of dynamical activity to current in both models, we identified the conditions under which the quantum signatures—such as reduced fluctuations and enhanced current stability—emerge. We also constructed equivalent classical models for both configurations, finding that quantum models consistently yield lower Fano factors and are thus more fluctuation-suppressed than their classical counterparts. Importantly, none of the classical models examined exhibited any KUR violations, highlighting the inherently quantum nature of the observed effects.

Our results demonstrate the relevance of quantum coherence and decoherence dynamics in determining the thermodynamic performance limits of quantum heat engines. Overall, our findings not only advance the theoretical foundations of KUR in open quantum systems but also highlight how decoherence pathways shape fluctuation behavior, thereby offering valuable insights for the design of quantum heat engines where stability and precision are also crucial.

 \section{Acknowledgements} 
This research was supported by individual KIAS Grants No. PG064902 (J.S.L.) and No. PG096801 (V.S.) at the Korea Institute for Advanced Study, and by the National Research Foundation of Korea (NRF) grants No. RS-2023-00278985 (E.K.) funded by the Ministry of Science and ICT (MSIT) of the Korea government.

\section{Data availability statement}
The data that support the findings of this study are available upon reasonable request from the authors.


\appendix\section{  Rate equations for three-level maser heat engine}
Here, we present the density matrix equations corresponding to the two distinct variants of the maser engine.
\subsection*{Model I}
For the three-level system depicted in Fig. 1(a), the time evolution of the density matrix elements is described by the following set of equations \cite{Dorfman2018,Varinder2020,BoukobzaTannor2007}:
\begin{eqnarray}
\dot{\rho}_{gg} &=&   \gamma_h(n_h+1)\rho_{11}+\gamma_c(n_c+1)\rho_{00}   - (\gamma_h n_h + \gamma_c n_c) \rho_{gg}, \nonumber
\\
\,
\\
\dot{\rho}_{11} &=& i\lambda (\rho_{10}-\rho_{01}) - \gamma_h[(n_h+1)\rho_{11}-n_h\rho_{gg}],\label{A1} \\
\dot{\rho}_{00} &=& -i\lambda (\rho_{10}-\rho_{01}) - \gamma_c[(n_c+1)\rho_{00}-n_c\rho_{gg}], \\
\dot{\rho}_{10} &=&  i\lambda(\rho_{11}-\rho_{00}) -\frac{1}{2}[\gamma_h(n_h+1)+\gamma_c(n_c+1)]\rho_{10}  , \nonumber
\\
\\
%
%
\dot{\rho}_{01} &=& \dot{\rho}_{10}^*. \label{A5}
\end{eqnarray}

The above equations can be solved in the steady-state regime, where ($\dot{\rho}_{ab}=0$), yielding the following solution:
\begin{eqnarray}
\rho_{10} &=&  \frac{2i(n_c-n_h)\gamma_h\gamma_c\lambda}
{
4\lambda^2[\gamma_h(3n_h+1)+\gamma_c(3n_c+1)] + (3n_h n_c+2n_h+2n_c+1)[\gamma_h(n_h+1)+\gamma_c(n_c+1)]\gamma_h\gamma_c
},  \label{cohI} \nonumber \\
\\
\rho_{gg} &=& \frac{\left[\gamma_c(n_c+1)+\gamma_h(n_h+1)\right]\left[4\lambda^2+(1+n_c)(1+n_h)\gamma_c\gamma_h\right]}
{
4\lambda^2[\gamma_h(3n_h+1)+\gamma_c(3n_c+1)] + (3n_h n_c+2n_h+2n_c+1)[\gamma_h(n_h+1)+\gamma_c(n_c+1)]\gamma_h\gamma_c
}, \label{rhoggI} \nonumber \\
\\
\rho_{00} &=& \frac{n_c(1+n_h)\left[\gamma_c(n_c+1)+\gamma_h(n_h+1)\right]\gamma_c\gamma_h+ 4\lambda^2(\gamma_c n_c+\gamma_h n_h)}
{
4\lambda^2[\gamma_h(3n_h+1)+\gamma_c(3n_c+1)] + (3n_h n_c+2n_h+2n_c+1)[\gamma_h(n_h+1)+\gamma_c(n_c+1)]\gamma_h\gamma_c
}, \label{rho00I} \nonumber \\
\\
\rho_{11} &=& \frac{n_h(1+n_c)\left[\gamma_c(n_c+1)+\gamma_h(n_h+1)\right]\gamma_c\gamma_h+ 4\lambda^2(\gamma_c n_c+\gamma_h n_h)}
{
4\lambda^2[\gamma_h(3n_h+1)+\gamma_c(3n_c+1)] + (3n_h n_c+2n_h+2n_c+1)[\gamma_h(n_h+1)+\gamma_c(n_c+1)]\gamma_h\gamma_c
}. \label{rho11I} \nonumber \\
\end{eqnarray}
\subsection*{Model II}
For the three-level system depicted in Fig. 1(b), the time evolution of the density matrix elements is governed by the following equations: %
 \begin{eqnarray}
\dot\rho_{11} &= &\gamma_h n_h\rho_{gg} +\gamma_c n_c\rho_{00}-[\gamma_h(n_h+1)+\gamma_c(n_c+1)] \rho_{11}, \nonumber
\\
\vspace{1mm} 
\\
\dot\rho_{00} &=& \gamma_c (n_c+1)\rho_{11} - \gamma_c n_c \rho_{00} + \imai \lambda (\rho_{0g}-\rho_{g0})\label{Eqrho00} ,
\\
\dot\rho_{gg} &=& \gamma_h (n_h+1)\rho_{11} - \gamma_h n_h\rho_{gg} -\imai\lambda (\rho_{0g}-\rho_{g0})\label{Eqrholl}\,,
\\
\dot\rho_{g0} &=&  \imai\lambda(\rho_{00}-\rho_{gg})-\frac{1}{2}(\gamma_h n_h+\gamma_c n_c) \rho_{g0}\,, 
\\
\dot\rho_{0g} &=& \dot\rho^*_{g0}  \label{Eqrhoul}
\end{eqnarray}

The steady state solution of the above equations is given by
 \begin{eqnarray}
   \rho_{g0} &=& \frac{2i(n_c-n_h)\gamma_h\gamma_c\lambda}
    {4\lambda^2[\gamma_h(3n_h+2)+\gamma_c(3n_c+2)] + (3n_h n_c+n_h+n_c)(\gamma_h n_h+\gamma_c n_c)\gamma_h\gamma_c
} , \label{cohII}
\\
   \rho_{gg} &=& \frac{n_c(1+n_h)(\gamma_c n_c+\gamma_h n_h) \gamma_h\gamma_c + 4\lambda^2\left[\gamma_c(1+n_c)+\gamma_h(1+n_h)\right]}
    {4\lambda^2[\gamma_h(3n_h+2)+\gamma_c(3n_c+2)] + (3n_h n_c+n_h+n_c)(\gamma_h n_h+\gamma_c n_c)\gamma_h\gamma_c
} , \label{rhoggII}
\\
   \rho_{00} &=& \frac{n_h(1+n_c)(\gamma_c n_c+\gamma_h n_h) \gamma_h\gamma_c + 4\lambda^2\left[\gamma_c(1+n_c)+\gamma_h(1+n_h)\right]}
    {4\lambda^2[\gamma_h(3n_h+2)+\gamma_c(3n_c+2)] + (3n_h n_c+n_h+n_c)(\gamma_h n_h+\gamma_c n_c)\gamma_h\gamma_c
} , \label{rho00II}
\\
   \rho_{11} &=& \frac{(\gamma_c n_c+\gamma_h n_h) \left[4\lambda^2+ n_c n_h\gamma_c\gamma_h\right]}
    {4\lambda^2[\gamma_h(3n_h+2)+\gamma_c(3n_c+2)] + (3n_h n_c+n_h+n_c)(\gamma_h n_h+\gamma_c n_c)\gamma_h\gamma_c
} . \label{rho11II}
\end{eqnarray}

\section{Full Counting Statistics}
\label{AppFCS}

To evaluate the average current $\langle I\rangle$ and variance $\Delta I$, we employ the formalism of full counting statistics (FCS), a powerful approach in open quantum systems that enables the characterization of particle transport by incorporating counting fields into the master equation. 
\subsection*{Model I}
In our analysis, it suffices to introduce a counting field for either the hot or cold reservoir. Without loss of generality, we choose to assign the counting field $\chi_c$  to the cold reservoir, following the approach outlined in Refs.~\cite{Bruderer2014,SchallerBook}.This results in a modified Lindblad master equation for Model I, incorporating the counting field, which takes the following form \cite{VSTUR2023}:
\begin{equation}
\dot{\rho} = - i [V_R,\rho] + \mathcal{L}_{h}[\rho] + \mathcal{L}^{\chi}_{c}[\rho],
\end{equation}
where $\mathcal{L}^{\chi}_{c}$, the modified Lindblad superoperator for Model I of maser heat engine, is given by
%
\begin{widetext}
\be
\mathcal{L}_c[\rho] = \gamma_c(n_c+1) \big (\e^{-i\chi} \sigma_{g0} \rho \sigma^{\dagger}_{g0} -\frac{1}{2}  \{\sigma^{\dagger}_{g0} \sigma_{g0},\rho\}   \big) +
\gamma_c n_c  \big (\e^{i\chi}  \sigma^{\dagger}_{g0} \rho \sigma_{g0} -\frac{1}{2}  \{\sigma_{g0} \sigma^{\dagger}_{g0} ,\rho\}   \big). \label{D2A}
\ee
Vectorizing the density matrix as $\rho_R=(\rho_{gg}, \rho_{00}, \rho_{11}, \rho_{10}, \rho_{01})^T$, the Lindblad master equation can be expressed as a linear matrix equation involving the Liouvillian supermatrix $\bm{\mathcal{ L}}(\chi_c)$
\begin{equation}
\dot{\rho} = \bm{\mathcal{ L}} (\chi_c) \rho,
\end{equation}
where
\end{widetext}
\begin{equation}
\setlength\arraycolsep{4pt} 
\renewcommand\arraystretch{1.08}
\resizebox{0.96\columnwidth}{!}{$
\bm{\mathcal{L}}^{\rm I}(\chi_c)=
\left[
\begin{matrix}
-(\gamma_h n_h + \gamma_c n_c) & \gamma_c (n_c+1)e^{-i \chi_c} & \gamma_h (n_h+1) & 0 & 0 \\
\gamma_c n_c e^{i\chi_c} & -\gamma_c(n_c+1) & 0 & -i\lambda & i\lambda \\
\gamma_h n_h & 0 & -\gamma_h(n_h+1) & i\lambda & -i\lambda \\
0 & -i\lambda & i\lambda & -\tfrac{1}{2}\!\left[\gamma_h(n_h+1)+\gamma_c(n_c+1)\right] & 0 \\
0 & i\lambda & -i\lambda & 0 & -\tfrac{1}{2}\!\left[\gamma_h(n_h+1)+\gamma_c(n_c+1)\right]
\end{matrix}
\right]
$}
\label{EqFCSLiouvillian1}
\end{equation}

In the asymptotic time limit, the $k$th  cumulant of the integrated photon count into the cold reservoir is determined by the following relation \cite{SchallerBook}:
\begin{equation}
C^k(t) =\left.(\imai\partial_{\chi_c})^k\left[\xi(\chi_c) t \right]\right|_{\chi_c=0} \equiv \left.(\imai\partial_{\chi_c})^k  \lambda^\prime(t) \right|_{\chi_c=0} , \label{cg1}
\end{equation}
where,  $\xi(\chi_c)$ denotes the eigenvalue of the tilted Liouvillian $\mathcal{L}(\chi_c)$ with the largest real part. In the long-time limit, the cumulant generating function for the integrated current is given by  $\lambda'(t)=\xi(\chi_c) t$   in Eq. (\ref{cg1}). To obtain the cumulants of the time-averaged current in this regime, we define the scaled cumulant generating function as follows \cite{SchallerBook, LandiReview2023}: 
\begin{equation}
\lambda'_{\rm scaled} = \lim_{t\rightarrow\infty} \frac{\lambda'(t)}{t}=\xi(\chi_c).
\end{equation} 
The first cumulant of the scaled cumulant generating function $\lambda'_{\rm scaled}$ corresponds to the mean current $\langle I\rangle$, , while the second cumulant yields the \textit{scaled} current variance, defined as $\Delta I=\lim_{t\rightarrow\infty}   \langle [I(t)  -\langle I\rangle]^2\rangle t$:
\begin{equation}
\langle I\rangle \simeq \left. \imai \partial_{\chi_c}\xi(\chi_c)\right|_{\chi_c = 0}, \quad
\Delta I \simeq\left. -\partial_{\chi_c}^2\xi(\chi_c)\right|_{\chi_c = 0}.
\label{EqTURCumulant}
\end{equation}
\begin{widetext}
To compute the mean current and its variance, we follow the method described in Ref.~\cite{Bruderer2014}, starting from the characteristic polynomial of the tilted Liouvillian $\mathcal{L}(\chi_c)$:
\begin{equation}
\sum_n a_n \xi^n = 0\,, \label{sum}
\end{equation}
where the coefficients are  $ a_n$ are functions of the counting field $\chi_c$. First and second derivatives of $a_n$ are defined as:
\begin{equation}
 a_n' = \imai\partial_{\chi_c}  a_n\Big\vert_{\chi_c=0},
\quad  a_n'' = (\imai\partial_{\chi_c})^2 a_n\Big\vert_{\chi_c=0}  .
\end{equation}
By taking the first order and second order derivatives of Eq.~(\ref{sum}) with respect to the counting parameter  $\chi_c$, and subsequently evaluating at  $\chi_c=0$, we obtain
\begin{equation}
\left[\imai\partial_{\chi_c}\sum_n  a_n\xi^n\right]_{\chi_c=0} = \sum_n[ a_n'+(n+1) a_{n+1}\xi']\xi^n(0) = 0\,,
\label{EqFCSfirst}
\end{equation}
\begin{equation}
\left[(\imai \partial_{\chi_c})^2\sum_n a_n\xi^n\right]_{\chi_c=0} =
\sum_n[ a_n''+2(n+1) a_{n+1}'\xi'+(n+1) a_{n+1}\xi''+(n+1)(n+2) a_{n+2}\xi'^2]\xi^n(0)=0\,.
\label{EqFCSsecond}
\end{equation}
Given that the steady state is associated with a zero eigenvalue of the Liouvillian,     $\xi^0=1$ should vanish, hence Eq.~\eqref{EqFCSfirst} yields the expression for current $\langle I\rangle = \xi'$:
\begin{equation}
 a_0'+ a_1\xi'=0 \quad \Rightarrow \langle I\rangle =\xi' = -\frac{ a_0'}{ a_1}.
\label{EqFCSmean}
\end{equation}
Similarly from Eq.~\eqref{EqFCSsecond},   following expression for the variance is obtained,
\begin{equation}
\Delta I = \xi'' =-\frac{ a_0''+2I( a_1'+ a_2I)}{ a_1} .
\label{EqFCSvariance}
\end{equation}

By applying the aforementioned method to the Liouvillian operator defined in Eq.~\eqref{EqFCSLiouvillian1}, we obtain following expressions for coefficients  $a_n$ and their   derivatives:

\begin{equation}
\begin{split}
 a_0'=& (n_h-n_c) \gamma_h\gamma_c   B \lambda^2,
\\
 a_0'' =&   (2n_h n_c+n_h+n_c)\gamma_h\gamma_c  B  \lambda^2,
\\
 a_1 =&  - \frac{B}{4}   \left(   \gamma_c\gamma_h   B\,C +  4 \lambda^2  D\right),
\\
 a_1' =& 2(n_h-n_c)\gamma_h\gamma_c\lambda^2,
\\
 a_2 =&  - \frac{B}{4} \Big\{  4 \lambda^2(B+D) +  4\gamma_c\gamma_h C + B \left[ (1+2n_c)\gamma_c + (1+2n_h)\gamma_h \right]    \Big\},
\end{split}
\end{equation}
where $B=\gamma_h (n_h+1)+\gamma_c (n_c+1)$, $C=3n_h n_c +2n_h+2n_c+1$, $D=\gamma_h(3n_h+1)+\gamma_c(3n_c+1)$. Using Eq. (\ref{EqFCSmean}), the expression for current is found to be
\begin{eqnarray}
\langle I^{\rm I}\rangle&=&  \frac{4(n_h-n_c)\gamma_h\gamma_c\lambda^2}
{
4\lambda^2[\gamma_h(3n_h+1)+\gamma_c(3n_c+1)] + (3n_h n_c+2n_h+2n_c+1)[\gamma_h(n_h+1)+\gamma_c(n_c+1)]\gamma_h\gamma_c 
}  \nonumber 
\\
&\equiv & \frac{4(n_h-n_c)\gamma_h\gamma_c\lambda^2}{4\lambda^2 D+B\,C\, \gamma_c\gamma_h}.
\label{CurrentI}
\end{eqnarray}
Similarly, expression for Fano factor, $F=\Delta I/\langle I \rangle$, can be written in the following simplified form:
\begin{equation}
    F^{\rm I} =  
    \frac{1}{n_h-n_c}\Bigg[
          2n_h n_c +n_h+n_c - \frac{8 \gamma_c\gamma_h\lambda^2 (n_h-n_c)^2 G}{(4\lambda^2 D+B\,C\, \gamma_c\gamma_h)^2}
\Bigg], \label{FanoI}
\end{equation}
where $G = 8 \lambda ^2+\gamma _c \gamma _h \left(  10 n_h n_c +7n_c +7 n_h+4\right)+\gamma _c^2 \left(n_c+1\right) \left(2 n_c+1\right)+\gamma _h^2 \left(n_h+1\right) \left(2 n_h+1\right)$.

In order to derive an analytical expression for the KUR ratio, it is essential to evaluate the dynamical activity as well. The dynamical activity for Model I is expressed as:
\begin{eqnarray}
    A^{\rm I} &=& \gamma_h(n_h+1) Tr\left[\sigma_{g1}\rho  \sigma^{\dagger}_{g1}\right] + \gamma_h n_h Tr\left[ \sigma^{\dagger}_{g1}\rho  \sigma_{g1}\right] +\gamma_c(n_c+1) Tr\left[\sigma_{g0}\rho \sigma^{\dagger}_{g0}\right] + \gamma_c n_c Tr\left[ \sigma^{\dagger}_{g0}\rho  \sigma_{g0}\right] \nonumber
    \\
    &=& (\gamma_h n_h +\gamma_c n_c)\rho_{gg} + \gamma_h (n_h+1) \rho_{11} +\gamma_c (n_c+1) \rho_{00}.
\end{eqnarray}
Plugging the expressions for $\rho_{11}$, $\rho_{gg}$ and $\rho_{00}$  from Eqs. (\ref{rhoggI})-(\ref{rho11I}), we get 
\begin{equation}
    A^{\rm I} = \frac{2(\gamma_c nc+\gamma_h n_h)[\gamma_c (1+nc)+\gamma_h (1+n_h)]\left[ 4\lambda^2+(1+n_c)(1+n_h)\gamma_c\gamma_h \right]}
    {4\lambda^2[\gamma_h(3n_h+1)+\gamma_c(3n_c+1)] + (3n_h n_c+2n_h+2n_c+1)[\gamma_h(n_h+1)+\gamma_c(n_c+1)]\gamma_h\gamma_c} \label{DAI}.
\end{equation}
Finally, using Eqs.~(\ref{CurrentI}), (\ref{FanoI}), and (\ref{DAI}), the KUR quantifier for the Model I is:
\begin{equation}
      \mathcal{Q}^{\rm I} = \frac{\langle I^{\rm I}\rangle^2}{A^{\rm I}\,F^{\rm I}} =\frac{\langle I^{\rm I}\rangle}{A^{\rm I}}\frac{1}{F^{\rm I}} =   \frac{2(n_h-n_c)^2\gamma_c\gamma_h\lambda^2}{B\,B'\,H }
\bigg[ 
        2n_h n_c +n_h+n_c - \frac{8\lambda^2(n_h-n_c)^2 \gamma_c\gamma_h G}{(4\lambda^2 D+B\,C\, \gamma_c\gamma_h)^2}
\bigg]^{-1}, \label{KURmeA}
\end{equation}
 where $B'=\gamma_c n_c+\gamma_h n_h$, $H=4\lambda^2 + (1+n_c)(1+n_h)\gamma_c\gamma_h$,
%
%
%
\subsection*{Model II}
By  repeating the  FCS method outlined above, an analytic expression for the KUR quantifier $\mathcal{Q^{\rm II}}$ is obtained. By transforming the density matrix into a state vector through vectorization (i.e., arranging its elements into a column vector) in the basis $\rho^{\rm II} = (\rho_{11}, \rho_{00},\rho_{gg}, \rho_{0g},\rho_{g0})^T$,  Liouvillian superoperator for Model II reads as
\begin{equation}
\bm{\mathcal{L}}^{\rm II} = \left[
\begin{matrix}
-[\gamma_h (n_h+1) + \gamma_c (n_c+1)] & \gamma_c n_c  & \gamma_h n_h & 0 & 0
\\
\gamma_c (n_c+1) & -\gamma_c n_c & 0 & i\lambda & -i\lambda
\\
\gamma_h (n_h+1)  & 0& -\gamma_h n_h & -i\lambda   & i\lambda 
\\
0 & i\lambda & -i\lambda & -\frac{1}{2}(\gamma_h n_h +\gamma_c n_c) & 0
\\
0 & -i\lambda & i\lambda & 0 &   -\frac{1}{2}(\gamma_h n_h +\gamma_c n_c) \\
\end{matrix}
\right]\,.
\label{EqFCSLiouvillian2}
\end{equation}
 Applying the FCS formalism to Model II, we obtain following closed-form expressions for the current, Fano factor, and dynamical activity:
\begin{equation}
    \langle I^{\rm II}\rangle = \frac{4(n_h-n_c)\gamma_h\gamma_c\lambda^2}
    {4\lambda^2[\gamma_h(3n_h+2)+\gamma_c(3n_c+2)] + (3n_h n_c+n_h+n_c)(\gamma_h n_h+\gamma_c n_c)\gamma_h\gamma_c
}  \equiv \frac{4(n_h-n_c)\gamma_h\gamma_c\lambda^2}{4\lambda^2 D' +B'\,C'\gamma_h\gamma_c}, \label{CurrentII}
\end{equation}
\begin{equation}
    F^{\rm II} =  
     \frac{1}{n_h-n_c}   \Bigg[2n_h n_c +n_h+n_c - \frac{8\lambda^2(n_h-n_c)^2 \gamma_c\gamma_h G'}{(4\lambda^2 D' +B'\,C'\gamma_h\gamma_c)^2 }
\Bigg], \label{FanoII}
\end{equation}
\begin{equation}
    A^{\rm II} = \frac{2(\gamma_c nc+\gamma_h n_h)[\gamma_c (1+nc)+\gamma_h (1+n_h)]\left( 4\lambda^2+n_c n_h \gamma_c\gamma_h \right)}
    {4\lambda^2[\gamma_h(3n_h+2)+\gamma_c(3n_c+2)] + (3n_h n_c+n_h+n_c)(\gamma_h n_h+\gamma_c n_c)\gamma_h\gamma_c
}\label{DAII},
\end{equation}
where $B'=\gamma_c nc+\gamma_h n_h$, $C'=3n_h n_c + n_h + n_c$, $D'=\gamma_c(2+3n_c)+\gamma_h(2+3n_h)$, $G' = 8 \lambda ^2+\gamma _c \gamma _h \left(  10 n_h n_c +3n_c +3 n_h\right)+\gamma _c^2 \left(2 n_c+1\right)n_c+\gamma _h^2   \left(2 n_h+1\right)n_h$. 
Using Eqs. (\ref{CurrentII})-(\ref{FanoII}), the final expression for KUR quantifier $\mathcal{Q}^{\rm II}$ for Model II is obtained as
\begin{equation}
\mathcal{Q}^{\rm II} =   \frac{2(n_h-n_c)^2\gamma_c\gamma_h\lambda^2}{B\,B'\,H' }  \Bigg[2n_h n_c +n_h+n_c - \frac{8\lambda^2(n_h-n_c)^2 \gamma_c\gamma_h G'}{(4\lambda^2 D' +B'\,C'\gamma_h\gamma_c)^2 }
\Bigg]^{-1},\label{KURPatrickA}
\end{equation}
where $H'=4\lambda^2 +  n_c n_h\gamma_c\gamma_h$.
 
 
 \section{ Equivalent Classical Models}
 To understand the difference between two different configurations, Model I and Model II of three-level maser heat engine, we construct classical equivalent models for both configurations. 
\subsection*{Model I}
 We replace the   coherent driving part $-\frac{i}{\hbar} [V_R^{\rm I},\rho]$  by incoherent part $\gamma_{\rm cl}^{\rm I} \left( \mathcal{D}_{\sigma_{10}}\left[\rho^{\rm I}_{\rm cl}\right]  +  \mathcal{D}_{\sigma_{01}}\left[\rho^{\rm I}_{\rm cl}\right]\right) $ in  Eq. (\ref{LindbladMain}), where $\gamma_{\rm cl}^{\rm I}$ is classical rate constant. The equivalent master equation  for Model I, consisting of jump operators only, is given by
\begin{equation}
    \dot{\rho}^{\rm I}_{\rm cl} =  \gamma^{\rm I}_{\rm cl} \left( \mathcal{D}_{\sigma_{10}}\left[\rho^{\rm I}_{\rm cl}\right]  +  \mathcal{D}_{\sigma_{01}}\left[\rho^{\rm I}_{\rm cl}\right]\right) + \mathcal{L}_{h}\left[\rho^{\rm I}_{\rm cl}\right] + \mathcal{L}_{c}\left[\rho^{\rm I}_{\rm cl}\right].\label{CMEA}
\end{equation} 
All coherences vanish within this formulation. Using the vectorized density matrix $\rho^{\rm I}_{\rm cl}=(\rho_{gg},\rho_{00},\rho{11})^T$, the Liouvillian  for the classical equivalent model reads as
\begin{equation}
\bm{\mathcal{L}}^{\rm I}_{\rm cl} = \left[
\begin{matrix}
-(\gamma_h n_h + \gamma_c n_c) & \gamma_c (n_c+1)  & \gamma_h (n_h+1) 
\\
\gamma_c n_c  & -\gamma_c(n_c+1)-\gamma_{\rm cl}^{\rm I}  & \gamma_{\rm cl}^{\rm I}
\\
\gamma_h n_h  & \gamma_{\rm cl}^{\rm I}  & -\gamma_h(n_h+1) -\gamma_{\rm cl}^{\rm I}
\\
\end{matrix}
\right]\,.
\label{EqFCSLiouvillianClassI}
\end{equation}
Applying FCS, the expression for current and Fano factor are obtained as follows:
\begin{equation}
    \langle I^{\rm I}_{\rm cl}\rangle = \frac{(n_h-n_c)\gamma_{\rm cl}^{\rm I}\gamma_c\gamma_h}
    {
\gamma_{\rm cl}^{\rm I}[\gamma_h(3n_h+1)+\gamma_c(3n_c+1)] + (3n_h n_c+2n_h+2n_c+1)\gamma_c\gamma_h} 
\equiv \frac{(n_h-n_c)\gamma_{\rm cl}^{\rm I}\gamma_c\gamma_h}{\gamma_{\rm cl}^{\rm I} D+C\gamma_c\gamma_h}
,
\label{CurrentClassI}
\end{equation}
 %
\begin{equation}
  F^{\rm I}_{\rm cl}  =\frac{2n_h n_c+n_h+n_c}{n_h-n_c}-\frac{2(n_h-n_c)\gamma_{\rm cl}^{\rm I}\gamma_c\gamma_h\left[\gamma_c(1+2n_c)+\gamma_h(1+2n_h)+2\gamma_{\rm cl}^{\rm I}\right]}   {  \left(\gamma_{\rm cl}^{\rm I} D+C\gamma_c\gamma_h\right)^2}. \label{FanoClassI}
\end{equation}

We next identify the conditions for which the currents in the two models coincide. Comparing the above equation with Eq.~(\ref{CurrentI}), we observe that the two currents coincide, $\langle I^{\rm I} \rangle = \langle I^{\rm I}_{\rm cl} \rangle$, when the coupling rate $\gamma{\rm cl}^{\rm I}$ is defined as
\begin{equation}
    \gamma_{\rm cl}^{\rm I} = \frac{4\lambda^2}{\gamma_h(1+n_h)+\gamma_c(1+n_c)}. 
\end{equation}
Finally, the dynamical activity for the classical model can be obtained by adding the contribution from incoherent transitions between the levels $\ket{0}$ and $\ket{1}$ to the dynamical activity of the quantum model, which is given by
\begin{eqnarray}
    A^{\rm I}_{\rm cl} =    A^{\rm I} + \gamma_{\rm cl}^{\rm I} \left(\rho_{11}+\rho_{00}\right).
\end{eqnarray}
\subsection*{Model II}
In Model II, to obtain the classical equivalent model, we replace  coherent driving part $-\frac{i}{\hbar} [V_R^{\rm II},\rho]$  by incoherent part $\gamma^{\rm II}_{\rm cl}\left( \mathcal{D}_{\sigma_{g0}}\left[\rho^{\rm II}_{\rm cl}\right]  +  \mathcal{D}_{\sigma_{0g}}\left[\rho^{\rm II}_{\rm cl}\right]\right) $ in Eq. (\ref{LindbladMain}). The equivalent master equation  for Model II, consisting of jump operators only, is given by
\begin{equation}
    \dot{\rho}^{\rm II}_{\rm cl} =  \gamma^{\rm II}_{\rm cl} \left( \mathcal{D}_{\sigma_{g0}}\left[\rho^{\rm II}_{\rm cl}\right]  +  \mathcal{D}_{\sigma_{0g}}\left[\rho^{\rm II}_{\rm cl}\right]\right) + \mathcal{L}_{h}\left[\rho^{\rm II}_{\rm cl}\right] + \mathcal{L}_{c}\left[\rho^{\rm II}_{\rm cl}\right].\label{CMEA}
\end{equation} 
 Using the vectorized density matrix $\rho^{\rm II}_{\rm cl}=(\rho_{11},\rho_{00},\rho{00})^T$, the Liouvillian  for the classical equivalent model can be wtitten as
\begin{equation}
\bm{\mathcal{L}}^{\rm II}_{\rm cl} = \left[
\begin{matrix}
-\left[\gamma_h (1+n_h) + \gamma_c (1+n_c)\right] & \gamma_c n_c)  & \gamma_h n_h 
\\
\gamma_c (1+n_c)  & -\gamma_c n_c-\gamma_{\rm cl}^{\rm II}  & \gamma_{\rm cl}^{\rm II}
\\
\gamma_h (1+n_h)  & \gamma_{\rm cl}^{\rm II}  & -\gamma_h n_h  -\gamma_{\rm cl}^{\rm II}
\\
\end{matrix}
\right]\,.
\label{EqFCSLiouvillianClassII}
\end{equation}
Applying FCS, the expression for current and Fano factor are obtained as follows:
\begin{equation}
    \langle I^{\rm II}_{\rm cl}\rangle = \frac{(n_h-n_c)\gamma_{\rm cl}^{\rm II}\gamma_c\gamma_h}
    {
\gamma_{\rm cl}^{\rm II}[\gamma_h(3n_h+2)+\gamma_c(3n_c+2)] + (3n_h n_c+n_h+n_c)\gamma_c\gamma_h}
\equiv \frac{(n_h-n_c)\gamma_{\rm cl}^{\rm II}\gamma_c\gamma_h}{ \gamma_{\rm cl}^{\rm II} D'+C'\gamma_c\gamma_h},
\label{CurrentClassII}
\end{equation}
\begin{eqnarray}
     F^{\rm II}_{\rm cl} =\frac{2n_h n_c+n_h+n_c}{n_h-n_c}-\frac{2(n_h-n_c)\gamma_{\rm cl}^{\rm II}\gamma_c\gamma_h\left[\gamma_c(1+2n_c)+\gamma_h(1+2n_h)+2\gamma_{\rm cl}^{\rm II}\right]}   {  \left(\gamma_{\rm cl}^{\rm II} D'+C'\gamma_c\gamma_h\right)^2}   
     \label{FanoClassII}
\end{eqnarray}
In Model II,  quantum and classical currents coincide, $\langle I^{\rm II} \rangle = \langle I^{\rm II}_{\rm cl} \rangle$, when we take  
\begin{equation}
    \gamma_{\rm cl}^{\rm II} = \frac{4\lambda^2}{\gamma_h n_h +\gamma_c n_c}. 
\end{equation}
As in Model I, the dynamical activity of the classical equivalent model is obtained by including the contribution from incoherent transitions between levels $\ket{g}$ and $\ket{0}$ in addition to the dynamical activity of the quantum model, and is given by
\begin{eqnarray}
    A^{\rm II}_{\rm cl} =    A^{\rm II} + \gamma_{\rm cl}^{\rm II} \left(\rho_{gg}+\rho_{00}\right).
\end{eqnarray}

\end{widetext}

 \end{document}